\documentclass{ws-procs9x6}
\def\ket#1{| #1 \rangle}

\def\comm#1#2{[ #1, #2 ]}
\begin{document}
\title{Bohm-Aharonov type effects in dissipative atomic systems\footnote{\uppercase{P}resented at \uppercase {DGMTP XXIII, N}ankai \uppercase{I}nstitute, \uppercase{T}ianjin: 25  \uppercase{A}ugust  2005
}}

\author{Allan I. Solomon}

\address{Department of Physics and Astronomy \\ 
The Open University, Milton Keynes MK7 6AA, UK\\ 
E-mail: a.i.solomon@open.ac.uk,}

\author{Sonia G. Schirmer}

\address{DAMTP, Cambridge University , UK \\
E-mail: sgs29@cam.ac.uk}  

\maketitle

\abstracts{
A state in quantum mechanics is defined as a positive operator of norm 1.  For finite systems, this may be thought of as a positive matrix of trace 1.  This constraint of positivity imposes severe restrictions on the allowed evolution of such a state.  From the mathematical viewpoint, we describe the two forms of standard dynamical equations - global (Kraus) and local (Lindblad) - and show how each of these gives rise to a semi-group description of the evolution.  We then look at specific examples from atomic systems, involving 3-level systems for simplicity, and show how these mathematical constraints give rise to non-intuitive physical phenomena, reminiscent of Bohm-Aharonov effects. 
  In particular, we show that for a multi-level atomic system it is generally impossible to isolate the levels, and this leads to observable effects on the population relaxation and decoherence. }

\section{Introduction}
The standard description of a quantum state suitable for an open system is by means of a density matrix $\rho$, a positive matrix of trace 1. For a {\em hamiltonian} (non-dissipative) 
system  one obtains a  unitary evolution of the state. 
 For a non-dissipative system
the time evolution of the density matrix ${\rho}(t)$ with
${\rho}(t_0)={\rho}_0$ is governed by
\begin{equation} \label{rhoevol}
  {\rho}(t) = {U}(t) {\rho}_0 {U}(t)^\dagger,
\end{equation}
where ${U}(t)$ is the time-evolution operator satisfying the
Schrodinger equation
\begin{equation} \label{eq:SE}
  i\hbar \frac{d}{dt}{U}(t) = {H}{U}(t), \qquad {U}(0)={I},
\end{equation}
where ${I}$ is the identity operator. The state ${\rho}(t)$ equivalently
satisfies the quantum Liouville equation
\begin{equation} \label{lse}
  i\hbar \frac{d}{dt}{\rho}(t) = [{H},{\rho}(t)]
  \equiv {H}{\rho}(t) - {\rho}(t){H}.
\end{equation}
${H}$ is the total Hamiltonian of the system.  (In the context of {\em Quantum Control} theory,  we may assume that $H\equiv {H}(\vec{f})$
depends on a set of control fields $f_m$:
\begin{equation} \label{eq:H_expansion}
  {H}(\vec{f}) = {H}_0 + \sum_{m=1}^M f_m(t) {H}_m,
\end{equation}
where ${H}_0$ is the internal Hamiltonian and ${H}_m$ is
the interaction Hamiltonian for the field $f_m$ for $1\le m\le M$.)
The advantage of the Liouville equation (\ref{lse}) over the
unitary evolution equation (\ref{rhoevol}) is that it can
easily be adapted for dissipative systems by adding a dissipation
(super-)operator ${L}_D[{\rho}(t)]$:
\begin{equation} \label{eq:dLE}
   i\hbar\dot{\rho}(t) = [{H},{\rho}(t)] + i\hbar {L}_D[{\rho}(t)].
\end{equation}

In general, uncontrollable interactions of the system with its
environment lead to two types of dissipation: phase decoherence
(dephasing) and population relaxation (decay). The former occurs
when the interaction with the enviroment destroys the phase
correlations between states, which leads to a decay of the
off-diagonal elements of the density matrix:
\begin{equation} \label{eq:dephasing}
 \dot{\rho}_{kn}(t)
  = -\frac{i}{\hbar}([{H},{\rho}(t)])_{kn}-\Gamma_{kn}\rho_{kn}(t)
\end{equation}
where $\Gamma_{kn}$ (for $k\neq n$) is the dephasing rate between
$\ket{k}$ and $\ket{n}$. The latter happens, for instance, when a
quantum particle in state $\ket{n}$ spontaneously emits a photon
and decays to another quantum state $\ket{k}$, which changes the
populations according to
\begin{equation} \label{eq:poptrans}
\dot{\rho}_{nn}(t) =
-\frac{i}{\hbar}([{H},{\rho}(t)])_{nn}
  +\sum_{k\neq n} \left[\gamma_{nk}\rho_{kk}(t)-\gamma_{kn}\rho_{nn}(t)\right]
\end{equation}
where $\gamma_{kn}\rho_{nn}$ is the population loss for level
$\ket{n}$ due to transitions $\ket{n}\rightarrow\ket{k}$, and
$\gamma_{nk}\rho_{kk}$ is the population gain caused by
transitions $\ket{k}\rightarrow\ket{n}$.  The population
relaxation rate $\gamma_{kn}$ is determined by the lifetime of
the state $\ket{n}$, and for multiple decay pathways, the
relative probability for the transition
$\ket{n}\rightarrow\ket{k}$.   Phase decoherence and population relaxation lead to a
dissipation superoperator (represented by an $N^2 \times N^2$
matrix) whose non-zero elements are
\begin{equation}
  \begin{array}{ll}
  ({L}_D)_{kn,kn} = -\Gamma_{kn} & k \neq n \\
  ({L}_D)_{nn,kk} = +\gamma_{nk} & k \neq n \\
  ({ L}_D)_{nn,nn} = - \sum_{n\neq k} \gamma_{kn}
 \end{array}
\end{equation}
where $\Gamma_{kn}$ and $\gamma_{kn}$ are positive numbers, with $\Gamma_{kn}$ symmetric in its indices.
The $N^2 \times N^2$ matrix superoperator $L_D$ may be thought of as acting on the $N^2$-vector $V$ obtained from $\rho$ by
\begin{equation}\label{vec}
  {V}_{[(i-1)N+j]}\equiv \rho_{ij}.
\end{equation}
The resulting vector equation is 
\begin{equation}\label{veceq}
  \dot{V}=L {V} = (L_H+L_D){V}
\end{equation}
where $L_H$ is the anti-hermitian matrix derived from the hamiltonian $H$.

The values of the relaxation and dephasing parameters may be determined by experiment, or simply chosen to supply a model for the dissipation phenomenon.   But they may not be chosen arbitrarily; the condition of positivity for the state $\rho$ imposes constraints on their values, as does their deduction from  rigorous theory.  We illustrate this by demonstrating  the constraint for a two-level system. 

 \section{Two-level systems}
 \subsection{Unitary evolution}
 The  general hamiltonian for a two-level system is given, up to an additive constant, by
 \begin{equation}\label{2ham}
H = w   \left[ \begin {array}{l@{\quad}r} 1 & 0 \\ 0 & -1 \end {array} \right] 
  + f_x \left[ \begin {array}{l@{\quad}r} 0 & 1 \\ 1 & 0 \end {array} \right] 
  + f_y \left[ \begin {array}{l@{\quad}r} 0 &-i \\ i & 0 \end {array} \right]
\end{equation}
which we assume here to be time-independent.

This corresponds to the (superoperator form) $L_H$, where
\begin{equation}\label{2sham}
L_H=\left[ \begin {array}{l@{\quad}c@{\quad}c@{\quad}r} 
0&i \left( {\it f_x}+i{\it f_y} \right) &i \left( -{\it f_x}+i{\it f_y} \right) &0\\\noalign{\medskip}-i \left( -{
\it f_x}+i{\it f_y} \right) &-2\,iw&0&i \left( -{\it f_x}+i{\it f_y}
 \right) \\\noalign{\medskip}-i \left( {\it f_x}+i{\it f_y} \right) &0&2
\,iw&i \left( {\it f_x}+i{\it f_y} \right) \\\noalign{\medskip}0&-i
 \left( {\it f_x}+i{\it f_y} \right) &-i \left( -{\it f_x}+i{\it f_y}
 \right) &0\end {array} \right]. 
\end{equation}
Note the useful rule for obtaining the equivalent $N^2\times N^2$ matrix
action $$A \rho B \Longleftrightarrow A \otimes B^{T} \; V.$$

The corresponding evolution equation for the  4-vector $V$ corresponding to the state $\rho$  is
\begin{equation}\label{hamv}
\dot{V}=L_H V.
\end{equation}
This is equivalent to  Eq.(\ref{rhoevol}), which clearly preserves the trace of $\rho$, and also its positivity, using the definition of a positive matrix as one of the form $MM^{\dagger}$. (Of course this result is true in general.)
\subsection{Pure dissipation}\label{pure}
The  dissipation (super-)operator is
 \begin{equation}\label{dissL}
  L_D= \left[ \begin {array}{cccc} -\gamma_{{21}}&0&0&\gamma_{{12}}
\\\noalign{\medskip}0&-{\it \Gamma}&0&0\\\noalign{\medskip}0&0&-{\it 
\Gamma}&0\\\noalign{\medskip}\gamma_{{21}}&0&0&-\gamma_{{12}}
\end {array} \right].
\end{equation}
The corresponding evolution equation
\begin{equation}\label{dissv}
\dot{V}=L_D V.
\end{equation}
has solution 
\begin{equation}\label{dissV}
V(t)=\exp(L_D t)V(0)
\end{equation}
which corresponds to a value of the state $\rho(t)$
\begin{equation} \label{dissrho}
  \left[ \begin {array}{cc} {\frac {\rho_{{11}}(\gamma_{{12}}+\gamma_{{21}}{E})+\gamma_{{12}}\rho_{{22}}(1-{E})}
  {\gamma_{{21}}+\gamma_{{12}
}}}&{e^{-t{\it \Gamma}}}\rho_{{12}}\\\noalign{\medskip}{e^{-t{\it 
\Gamma}}}\rho_{{21}}&{\frac {\gamma_{{21}}\rho_{{11}}(1-{E})+
\rho_{{22}}(\gamma_{{21}}+\gamma_{{12}}E ) }{\gamma_{{21}}+\gamma_{{12}}}}
\end {array} \right] 
\end{equation}
where $E=e^{-t \left( \gamma_{{21}}+\gamma_{{12}} \right)}$,
for which it may readily be checked that ${\rm Tr} \rho(t)=\rho_{11}+\rho_{22}=1$.
Additionally, $\det\rho(t)$ is given by 
\begin{equation}\label{detrho}
\rho_{{11}}\rho_{{22}}{e^{-t \left( \gamma_{{21}}+\gamma_{{12}}
 \right) }}- \left( {e^{-2t{\it \Gamma}}} \right) \rho_{{12}}\rho_{
{21}}+2\,{\frac {\rho_{{11}}\gamma_{{12}}\rho_{{22}}\gamma_{{21}}
 \left( 1-{e^{-t \left( \gamma_{{21}}+\gamma_{{12}} \right) }}
 \right) ^{2}}{ \left( \gamma_{{21}}+\gamma_{{12}} \right) ^{2}}}
\end{equation}
which is clearly positive for all $t$ when
\begin{equation}\label{cond}
 2\Gamma \geq \gamma_{12}+\gamma_{21}
\end{equation}
since 
$\det \rho(t) \geq e^{-t \left( \gamma_{21}+\gamma_{12}\right)} \det \rho(0)\geq 0$.
Conversely, when the condition Eq.(\ref{cond}) is violated,  it is easy to display examples for which the evolution does not produce a state. 
For example, for a pure state, which satisfies $\rho_{11}\rho_{22}-\rho_{12}\rho_{21}=0$, choosing $\gamma_{12}>2\Gamma, \gamma_{21}=0$, Eq.(\ref{detrho}) is clearly negative.
\subsection{General dissipation}
When the hamiltonian matrix $L_H$ and the dissipation matrix $L_D$ commute, the conclusions of the  previous two subsections produce the same constraint for the solution of Eq.(\ref{veceq}).  In the general case   these matrices do not commute; they do however generate a local semi-direct group. More accurately the Lie algebra is locally a semi-direct sum\footnote{In the present two-level case, the local Lie algebra is the 12-element $gl(3,R)\oplus R^3$, and  in general $gl(N^2-1,R)\oplus R^{N^2-1}$, as discussed in\cite{paris}.}, which then generates a {\em semi-group}. In this case also, general theory, which we discuss in the next section, shows that the trace and determinant conditions of Eq.(\ref{cond}) remain  unchanged.
\section{Rigorous formulations}
\subsection{Kraus formalism and semi-groups}
The global form of the evolution equation Eq.(\ref{rhoevol}) in the presence of dissipation is due to Kraus\cite{Kraus}. The evolution of the state $\rho$ is given by 
\begin{equation} \label{Krausevol}
  {\rho}(t) = \sum_i{{W}_i(t) {\rho}_0 {W}_i(t)^\dagger},
\end{equation}
with 
\begin{equation}\label{Krauscond}
\sum_i W_i(t)^\dagger W_i(t) = I.
\end{equation}
Equation(\ref{Krausevol}) and the condition Eq.(\ref{Krauscond}) clearly guarantee both positivity and unit  trace. 

Further, though less obviously, this system implies the existence of a {\em semi-group} description of the evolution. For if we consider the set $G$ whose elements are the {\em sets} $\{w_i\}$ satisfying Eq.(\ref{Krauscond}), then if $g=\{w_i\}$ and $g'=\{w'_i\}$ are two elements of $G$, then so too is $g  g'$, where the product is taken in the sense of set multiplication. Although closed under composition, the only elements of $G$ which possess inverses are the singleton sets $\{U \}$, where $U$ is unitary.
\subsection{Lindblad formalism}
In so far as the Kraus formalism provides an analogue of the unitary evolution equation Eq.(\ref{rhoevol}), the Lindblad\cite{Lindblad} formalism gives an analogue of the Schroedinger equation Eq.(\ref{lse}):
\begin{eqnarray}\label{lindblad}
\dot{\rho}(t)&=&L[{\rho}(t)] {\rho(t)} \nonumber \\
&=& -i[H,\rho(t)] 
+ \frac{1}{2}\sum_k \left( \comm{{V}_k {\rho}(t)}{{V}_k^\dagger}+
                      \comm{{V}_k}{{\rho}(t){V}_k^\dagger} \right) 
\end{eqnarray}
where the $V_k$ are $N\times N$ matrices, but otherwise arbitrary\footnote{We may also choose an arbitrary number of  matrices $V_k$.}.
It may be proved that the dissipation superoperator $L_D$ arising from Eq.(\ref{lindblad}) has negative eigenvalues. Since the evolution dynamics arises from exponentiation of  $L_D t$ it follows that operators $\exp(L_D t)$ in the theory will become unbounded for arbitrary negative $t$. This means that not all operators will have inverses and implies a semi-group character to the evolution, as in the Kraus formalism.  
\subsection{$2\times 2$ Lindblad example}

Choosing four independent complex $V$-matrices 
$$V_{{1}}= \left[ \begin {array}{l@{\quad}r} 
a_{{1}}&0\\\noalign{\medskip}0&0
\end {array} \right] \; \; \; 
V_{{2}}= \left[ \begin {array}{l@{\quad}r}  0&a_{{2}}\\\noalign{\medskip}0&0
\end {array} \right] \; \; \; 
V_{{3}}= \left[ \begin {array}{l@{\quad}r}  0&0\\\noalign{\medskip}a_{{3}}&0
\end {array} \right] 
 \; \; \; 
V_{{4}}= \left[ \begin {array}{l@{\quad}r}  0&0\\\noalign{\medskip}0&a_{{4}}
\end {array} \right] 
$$
we obtain for the dissipation superoperator $L_D$
$$
 \left[ \begin {array}{cccc} 
  -|a_3|^2&0&0& |a_2|^2 \\\noalign{\medskip}
  0&-1/2\,A &0&0 \\\noalign{\medskip}
  0&0&-1/2\,A &0 \\\noalign{\medskip}    
  |a_3|^2&0&0&-|a_2|^2
\end {array} \right].
 $$
where $A=|a_1|^2+|a_2|^2+|a_3|^2+|a_4|^2$, which on comparison with Eq.(\ref{dissL}) gives,
 defining $\tilde{\Gamma}\equiv \frac{1}{2}(\left| a_{1} \right|^2+\left| a_{4} \right|^2)$

 $$
 \gamma_{21}=\left| a_3 \right|^2,\; \;
 \gamma_{12}=\left| a_2 \right|^2,\; \;
 \Gamma=\tilde{\Gamma}+\frac{1}{2}(\gamma_{12}+\gamma_{21})
$$
whence the constraint Eq.(\ref{cond}).  Note that $(\gamma_{12}+\gamma_{21})/2$ is the phase decoherence forced by population relaxation and $\tilde{\Gamma}$ is the contribution of pure dephasing.
\subsection{General $N \times N$ Lindblad case}
A convenient choice for the $V_k$ matrices may be made by defining
$$V_{[i,j]}=a_{[i,j]} E_{ij} $$
where $E_{ij}$ is the standard basis for $N\times N $ matrices, with $(E_{ij})_{\alpha \beta}=\delta_{ia}\delta_{j\beta}$ and we use the  index notation $[i,j] \equiv (i-1)N+j$.
The relaxation and decoherence parameters are defined by
\begin{eqnarray}\label{pars}
\gamma_{ij}&=&\left| a_{[i,j]}\right| ^{2} \; \; \; (i\neq j)\nonumber \\
\tilde{\Gamma}_{ij}&=&\frac{1}{2}(\left| a_{[i,i]}\right| ^{2}+\left| a_{[j,j]}\right| ^{2} ) \; \; \; (i\neq j)\nonumber \\
\Gamma_{ij}&=&\frac{1}{2}\sum_k(\left| a_{[k,i]}\right| ^{2}+\left| a_{[k,j]}\right| ^{2} ) \; \; \; (i\neq j)
\end{eqnarray}
\section{Bohm-Aharonov type effects}
What we mean by {\em Bohm-Aharonov type effects} in  the title of this note, and of this section, is the impossibility of isolation of quantum subsystems. We illustrate this type of effect by considering the use  of a two-level atomic system as, say, a qubit, when this is a subsystem of a multi-level system. 

We consider the case of pure dissipation as discussed in subsection \ref{pure}. Choosing values $\gamma_{21}=0,\; \; \gamma_{12}=\gamma,\; \; \Gamma=\frac{1}{2}\gamma$, which satisfy the constraint Eq.(\ref{cond}),  the state evolution is given by  
\begin{equation}\label{twolev}
\rho(t) =
\left[ \begin {array}{l@{\quad}r} 
\rho_{{11}}+\rho_{{22}} \left( 1-{e^{-t\gamma}} \right) &{e^{-1/2\,t\gamma}}\rho_{{12}}\\\noalign{\medskip}{
e^{-1/2\,t\gamma}}\rho_{{21}}&\rho_{{22}}{e^{-t\gamma}}\end {array} \right]
\end{equation}
where the initial state is 
$$\rho(0)= \left[ \begin {array}{cc} \rho_{{11}}&\rho_{{12}}
\\\noalign{\medskip}\rho_{{21}}&\rho_{{22}}\end {array} \right] $$

We now assume that our two-level system is embedded in a three level system, so that the state's evolution is given by 
\begin{equation}\label{threelev}
\rho(t)=
 \left[ \begin {array}{l@{\quad}c@{\quad}r} 
\rho_{{11}}+\rho_{{22}} \left( 1-{e^{-t
\gamma}} \right) &{e^{-1/2\,t\gamma}}\rho_{{12}}&\rho_{{13}}
\\\noalign{\medskip}{e^{-1/2\,t\gamma}}\rho_{{21}}&\rho_{{22}}{e^{-t\gamma}}&\rho_{{23}}\\\noalign{\medskip}
\rho_{{31}}&\rho_{{32}}&\rho_{{33}}\end {array} \right].
\end{equation}
Now consider three examples for the state evolution.  In all cases we start off with a pure state, in the first case with the third level not being populated. 
\subsection{Unpopulated third level}\label{utl}
Assume an initial pure state represented by the $3$-vector $v=[1/\sqrt {2},1/\sqrt {2},0]$ corresponding 
to the density matrix
\[ \frac{1}{2} 
\left[ \begin {array}{l@{\quad}c@{\quad}r} 1 & 1 & 0\\ 1 & 1 & 0\\ 0 & 0 & 0 \end {array} \right].
\]
Assuming that the third level is unaffected, the state evolution is given by (measuring $t$ in units of $1/\gamma$)
\begin{equation}\label{pure1}
\rho(t)=
 \left[ \begin {array}{l@{\quad}c@{\quad}r} 
1-1/2\,{e^{-t}}&1/2\,{e^{-1/2\,t}}&0\\\noalign{\medskip}1/2\,{e^{-1/2\,t}}&1/2\,{e^{-t}}&0
\\\noalign{\medskip}0&0&0\end {array} \right]. 
\end{equation}
In this case the naive picture of the evolution is justified, as the third level remains unpopulated, the eigenvalues remain positive ($\geq 0$), and the extra levels are not affected by the dissipative dynamics. The third level plays no role in the evolution.
However, in general an upper level will not be totally unpopulated; and in this case the constraints play a role.
\subsection{Equally populated third level}
We take the initial pure state vector to be 
\begin{equation}\label{eqpop}
 v=[1/\sqrt {3},1/\sqrt {3},1/\sqrt {3}]
\end{equation}
 giving the evolution
\begin{equation}\label{pure2}
\rho(t)= \frac{1}{3}
 \left[ \begin {array}{l@{\quad}c@{\quad}r} 
   2-{e^{-t}}&{e^{-1/2\,t}}&1\\\noalign{\medskip}
   {e^{-1/2\,t}}&{e^{-t}}&1\\\noalign{\medskip}
   1&1&1
\end {array} \right] . 
\end{equation}
As in subsection \ref{utl} we have assumed that the third levels are {\em not} affected by the dissipative dynamics. However, a numerical calculation shows that  the eigenvalues of $\rho(t)$  are not all positive; therefore the {\it assumed} evolution does not give a state, and so the naive assumption that the other levels remain unaffected is {\it false}. 
\subsection{Pure dephasing}
Population relaxation is not the only source of constraints on the 
decoherence rates for $N>2$.  Even
if there is no population relaxation at all, i.e., $\gamma_{kn}=0$ for all $k,n$, 
and the system experiences only pure dephasing, we cannot choose the decoherence 
rates $\Gamma_{kn}$ arbitrarily.  For example, setting $\Gamma_{12}\neq 0$ and 
$\Gamma_{23}=\Gamma_{13}=0$ for our three-level system gives
\begin{equation} \label{eq:false2}
  {\rho}(t) =
  \left[ \begin{array}{ccc}
   \rho_{11} & e^{-\Gamma_{12} t} \rho_{12} & \rho_{13} \\
   e^{-\Gamma_{12} t} \rho_{21} & \rho_{22} & \rho_{23} \\
   \rho_{31} & \rho_{32} & \rho_{33}
  \end{array} \right]   .
\end{equation}
Choosing ${\rho}(0)$ as in Eq.~(\ref{eqpop}) we again obtain a density operator
${\rho}(t)$ with negative eigenvalues, as a simple  calculation will reveal.  This shows that 
there must be additional constraints on the decoherence rates to ensure that the 
state of the system remains physical.
\section {Conclusions}
We have shown that it is impossible to isolate   a two-level system from  a multi-level system in the sense of  assuming that the other levels will not be affected by relaxation and decoherence effects in the ``isolated'' system.  A more general treatment of the effects noted here may be found elsewhere\cite{schsol}; in that paper the constraints are explicitly described for some multilevel systems, and  the effects of these constraints   are discussed.

\end{document}